\documentstyle[epsf]{mn}

\def\arcsec{\hbox{$^{\prime\prime}$}}
\newcommand{\kms}{\hbox{\textrm km\,s$^{-1}$}}
\newcommand{\um}{\hbox{$\mu$m}}
\newcommand{\FeII}{[Fe\,II]}

\newcommand{\PaB}{\hbox{Pa$\beta$}}
\newcommand{\OI}{[O\,I]}

\newcommand{\OIII}{[O\,III]}
\newcommand{\SIX}{[S\,IX]}
\newcommand{\HII}{H\,II}
\newcommand{\HeI}{He\,I}
\title{The origin of \FeII\ emission in NGC~4151}
\author[J.E.H.\ Turner et al.]
{James E.H.\ Turner$^1$\thanks{Now at: Gemini Observatory, Casilla 603, La
Serena, Chile.},
Jeremy Allington-Smith$^1$\thanks{Corresponding author:
  j.r.allington-smith@durham.ac.uk} 
Scott Chapman$^2$\thanks{Now at: California Institute of Technology, 1200
East California Boulevard, Pasadena, CA 91125, USA.},
Robert Content$^1$,
\newauthor
Christine Done$^1$, 
Roger Haynes$^1$\thanks{Now at: Anglo-Australian Observatory, P.O.\ Box
296, Epping, NSW 1710, Australia.}, 
David Lee$^1$\thanks{Now at: UK Astronomy Technology Centre, Royal
Observatory Edinburgh, Blackford Hill, Edinburgh EH9 3HJ}  and
Simon Morris$^2$\thanks{Now at: Department of Physics, University of
Durham, South Road, Durham DH1 3LE, UK.} \\
$^1$ University of Durham, Physics Department, South Rd, Durham DH1 3LE, UK\\
$^2$ Herzberg Institute of Astrophysics, 5071 West Saanich Road, Victoria, BC, Canada, V9E 2E7
}

\date{Accepted ;      Received ;       in original form} 
 
\pagerange{\pageref{firstpage}--\pageref{lastpage}} 

\pubyear{} 

\begin{document}

\maketitle

\begin{abstract}
The centre of NGC\,4151 has been observed in the J-band with the SMIRFS
integral field unit (IFU) on the UK Infrared Telescope. A map of \FeII\
emission is derived, and compared with the distributions of the optical
narrow line region and radio jet. We conclude that, because the \FeII\
emission is associated more closely with the visible narrow-line region than
with the radio jet, it arises mainly through photoionization of gas by
collimated X-rays from the Seyfert nucleus.  The velocity field and strength
with respect to \PaB\ are found to be consistent with this argument. The
performance of the IFU is considered briefly, and techniques for observation
and data analysis are discussed.
\end{abstract}

\section{Introduction}

Seyfert galaxies exhibit strong emission lines from forbidden electronic
transitions---notably those of oxygen, nitrogen and sulphur at visible
wavelengths \cite{Seyfert43} and iron in the near-infrared
\cite{Rieke81,Simpson96a}. These transitions can occur where ions are
excited by collisions with electrons, yet can decay radiatively before
subsequent collisions lead to de-excitation \cite{Peterson97,Osterbrock89}.
Hence the lines are associated with low-density regions of ionized hydrogen,
where free electrons are abundant but the timescale for collisions is
sufficiently long.

Some species, including neutral oxygen and Fe$^+$, have similar ionization
potentials to hydrogen, so do not survive where it is fully ionized. The
most favourable conditions for \OI\ and \FeII\ emission arise in partially
ionized zones, the extent of which influences the observed intensity. Since
\HII\ regions around stars normally give way sharply to neutral gas beyond
the Str\"omgren radius, their \OI\ and \FeII\ lines are relatively weak. The
central spectra of Seyferts, in contrast, indicate the presence of extensive
partially-ionized hydrogen clouds as well as very highly-ionized
species. These differences are explained somewhat naturally if the emission
line clouds in Seyferts are photoionized by the `power-law' continuum of the
central active galactic nucleus (AGN), which is spectrally flatter than a
black body \cite{Osterbrock89,Mouri90}. Numerical simulations
\cite{Ferland83,Mouri00} lend credence to this scenario.

The second mechanism which can generate enhanced zones of \OI\ and \FeII\
emission is shock excitation of gas \cite{Osterbrock89}. This appears to be
the dominant cause in starburst galaxies, where shocks due to supernova
remnants are important and supernovae have been associated directly with
enhanced \FeII\ \cite{Lester90,Greenhouse91,Forbes93a,Greenhouse97}.
Comparison of observed spectral line ratios with recent numerical models
helps to distinguish between shocking and photoionization \cite{Mouri00},
and indeed proves consistent with the former dominating in starbursts and
the latter in Seyfert nuclei. Doppler broadening can also be used to check
for kinematic disturbances.

Long-slit spectroscopy of Seyfert galaxies has revealed a spatially extended
narrow line region (ENLR) in the visible (at radius greater than 4 arcsec in
the case of NGC 4151) beyond the central narrow-line region (NLR).  Unger et
al. (1987) traced the bright \OIII\ lines at 4960\AA\ and 5007\AA, generated
in fully-ionized regions, along with H$\beta$ (4861\AA) in a sample of
galaxies. The features were shown to have minimal line-of-sight velocity
structure consistent with simple rotation, and line ratios characteristic of
high excitation levels, pointing once again to photoionization. Subsequent
studies \cite{Penston90,Robinson94} have tended to support this
conclusion. The elongated, sometimes visibly conical morphology of ENLR (and
NLR resolved with the Hubble Space Telescope: HST) points to a link with
collimated radiation from the nucleus \cite{Unger87,Mulchaey96,Evans93}.

In the near-infrared, the origin of strong, extended \FeII\ emission in
Seyfert galaxies has been somewhat controversial, with evidence for both
shock excitation and photoionization occurring. It has been argued
\cite{Greenhouse91} that shocks must be important because iron in our
interstellar medium is mainly condensed onto grains which shocks can destroy
via sputtering \cite{Savage96}. However, recent studies
\cite{Simpson96a,Mouri00} indicate that the ratio \FeII/\OI\ in both
Seyferts and starbursts is consistent with the usual depletion levels of
iron and oxygen from the gaseous phase, and that certain other signs of
grain destruction are absent. Even if metal enhancement is not the dominant
factor, it is likely that shocks due to outflows along the radio jets of AGN
do make some contribution to emission from partially-ionized zones
\cite{Simpson96a,Morse96}. In particular, Knop et al. 1996 find that whilst
\FeII\ in NGC\,4151 seems kinematically compatible with the ENLR, the
1.257\,$\mu m$ feature is broader than the nearby Pa$\beta$ line

NGC\,4151 is both a natural candidate in which to study \FeII\ excitation
and an important test case for understanding the structure of active
galaxies. It is one of the brightest nearby Seyfert galaxies
($\mathrm{V}\simeq11.5$, $cz=995$\,\kms), yet eludes firm classification in
the framework of unified AGN models \cite{Antonucci93}. The highly extended
narrow line emission and radio jet \cite{Perez89,Pedlar93} are
characteristic of a Seyfert 2 nucleus, whose collimation axis is almost
perpendicular to the line of sight. However, the object also exhibits the
strong broad lines and variability of a Seyfert 1 \cite{Maoz91}, suggesting
quite a different orientation. NGC\,4151 has therefore been the subject of
many past projects involving slit spectroscopy and narrow-band imaging.

An unusual feature which facilitates the present investigation is the
pronounced misalignment between the ENLR at position angle $\sim$50$^{\circ}$
\cite{Perez89,Evans93,Kaiser00}, and radio jet at $77^{\circ}$
\cite{Pedlar93,Mundell95}. Whatever its origin, this provides an
opportunity to associate \FeII\ directly with the optical ENLR and/or radio
jet, by tracing the emission along both axes. If a clear separation can be
made, it will be possible to determine the relative influence of ionization
by collimated X-rays and shocking by outflowing radio plasma. Such a task
is inherently suited to integral field spectroscopy, which can form a
complete, homogeneous picture without prior assumptions, slit alignment
problems or the waveband limitations of a filter.

The intrumentation is discussed in the next section, followed, in Section 3,
by details of the observations. In Section 4, we present the method of data
reduction and analysis which presents particular challenges since the
technological limitations required the fibre outputs to overlap at the slit.
However this does not degrade the data provided that the field is critically
sampled at the IFU input \cite{AC98}. Finally, the results are discussed in
Section 5 and compared with other work.

\section{Instrumentation for integral field spectroscopy}

The SMIRFS-IFU is an experimental 72-fibre integral field unit
\cite{SMIRFS99,LeePhD} feeding the CGS4 spectrograph\cite{Wright93} at the
UK Infrared Telescope (UKIRT). It provides integral field spectroscopy in
the J and H bands via lensed fused-silica fibres.  It was designed and built
by the Astronomical Instrumentation Group at Durham University as a
prototype for much larger integral field units: TEIFU on the William
Herschel Telescope\cite{Teifu00} and the GMOS-IFU on Gemini
\cite{GMOS00}. Futher details may be found in Haynes et al. (1999).

SMIRFS is designed to intercept the telescope beam without modification to
existing instrumentation, but consequently requires removal of the CGS4
calibration unit. In place of this module, the SMIRFS slit projection unit
feeds light from the fibre bundle into the long-slit, circumventing the
usual beam path; the IFU input is connected to another port of the UKIRT
instrument support structure. As a result of this arrangement, any
reference observations must be taken using the main telescope aperture.

At the image plane, a hexagonal microlens array provides optimal coupling to
the slow telescope beam and avoids dead space between fibres (see
Haynes et al. for more information). The field of view is $6 \times 4$
arcsec, with 0.62 arcsec spacing between adjacent fibre centres. A second
set of microlenses at the output restore the correct focal ratio for
CGS4. When using the long camera, each fibre projects a spot whose FWHM is
$\sim$1.8 pixels at the detector, giving a comparable spectral resolution to
the standard 2 pixel wide slit. These spots are spaced two pixels apart,
placing the instrument on the borderline between the `maximally packed' and
`resolved peaks' categories defined by Turner (2001), although
throughput variations obscure the regular pattern in practice. Since there
is only one field, sky subtraction is performed by nodding the telescope off
source.

The capability of the SMIRFS-IFU was previously demonstrated in a June 1997
commissioning run. Its throughput relative to CGS4 alone was measured to be
$\sim$50\%; since then, minor adjustments have been made to improve
performance. 

\section{Observations}

\subsection{Science programme}

NGC\,4151 was observed with the SMIRFS-IFU on 15--17 February 1998, as part
of a sample of Seyfert galaxies.  Details of the measurement parameters are
presented in table~\ref{N4151/observ}.  The selected region of the
J-band includes both \FeII\ and \PaB, at rest wavelengths of 1.2567\um\ and
1.2818\um\ respectively. Previous studies have sometimes used \FeII\ at
1.6435\um\ and Br$\gamma$ at 2.1655\um; such measurements are equivalent to
ours, because the same upper energy level of Fe$^+$ is involved. Since,
however, the longer-wavelength lines are much further apart, they cannot be
observed together at high dispersion and their ratio is more sensitive to
differential extinction along the line of sight. The J-band \FeII\ feature
is also brighter, and the uncooled SMIRFS better suited to working in that
regime.

\setcounter{table}{0}
\begin{table}
\label{N4151/observ}
\caption{Details of NGC\,4151 observations}
\begin{tabular}{ll}
\hline
  Instrument             & CGS4 + SMIRFS-IFU             \\
  Camera                 & 300mm                         \\
  Grating                & 150 lines/mm                  \\
  Diffraction order      & 3                             \\
  Filter                 & B1                            \\
  Spectral range         & 1.25--1.31\um                 \\
  Spectral FWHM          & $\sim 4.2$\AA$\ (100\,\kms)$  \\
  Spectral sampling      & 2.1\AA                        \\
%\hline
  IFU field        & $6\times 4$\arcsec                   \\
  Spatial FWHM     & $\sim 1.2\times 1.1$\arcsec           \\
  Spatial sampling & $0.62\times 0.54$\arcsec              \\
  Detector         & $256\times 256$ InSb                  \\
  Gain             & 6\,e$^-$/data number                  \\
  Read noise       & 23\,e$^-$                             \\
  Dark current     & $\sim 1\,$e$^{-}\!$/s 
%($\sigma=6\,$e$^{-}$)
                                                           \\
  Single exposure  & 180s $\times$ 2 detector positions    \\
  Central exp.     & 48 minutes on source                  \\
\hline
\end{tabular}
\end{table}

%http://www.jach.hawaii.edu/JACpublic/UKIRT/instruments/cgs4/cgs4.html
%http://www.stats.uwo.ca/courses/ss241b1/normal_table.html

The spatial resolution quoted in table~\ref{N4151/observ} is for the
final, combined data set; it was estimated by producing images of the
unresolved \SIX\ line and \PaB\ broad line. These FWHM values correspond to
the Nyquist scale, suggesting that the seeing disc was
undersampled---the implications are discussed briefly in
\S\ref{s:N4151/dred}.

The long axis of the IFU was orientated at $50^{\circ}$ east of north, along
the optical ENLR of the galaxy. The field of view, however, only covers the
inner NLR at $<4\arcsec$ \cite{Winge99}, which is orientated less favourably
at $\sim$60$^{\circ}$ \cite{Kaiser00}. With arcsecond resolution, this gives
$\sim$1\arcsec\ FWHM separation from the radio axis. Hence a mosaic was
made of several telescope pointings around the nucleus.

Target centering was performed with the aid of a dedicated program, which
derives a rough map of the illuminated input elements from the data. For
the peripheral mosaic positions, it was necessary to rely on specified
telescope offsets from the centre. Each position was observed in turn,
following the repeated sequence object--sky--object; this maximizes the
time on source without giving unequal object and sky exposures at
individual pointings. The 180s integration time was just long enough for
the background to dominate the read noise and dark current. The whole
mosaic was repeated until the full exposure was reached.

In the absence of the calibration unit, sky emission lines were used as a
wavelength reference. Some arc lamp observations were also taken through
the telescope. The IFU was flat-fielded using the illuminated dome and the
twilight sky, whilst a detector flat was made with the calibration unit,
prior to the installation of SMIRFS. The bright F and A stars BS4572 and
HD105601 were observed in order to calibrate the spectral response of the
instrument and measure sky absorption features.

\subsection{IFU set-up and characteristics}

\begin{figure}
%\vspace{10cm}
\epsfysize 9 truecm
\epsffile{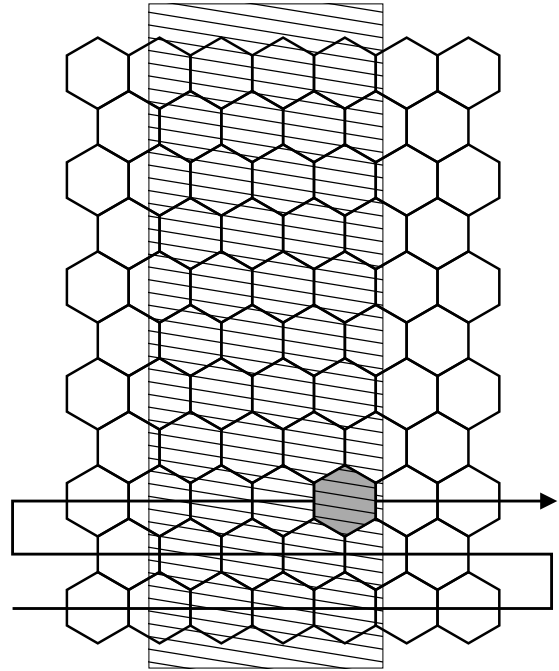}
\caption
{Sketch of the SMIRFS-IFU input with a mask in place for determining the
output magnification. The line with an arrow indicates the fibre ordering at
the slit. The dead fibre is shaded.}
\label{N4151/ifuinput}
\end{figure}

The IFU was aligned at the focal plane by projecting a laser beam back
through the fibres onto the UKIRT secondary mirror. The CGS4 slit was set
parallel to the row of fibre outputs, before adjusting the optics of the
slit projection unit so that fibres were imaged two pixels apart in the
spectrograph. The two pixel magnification was verified by illuminating the
IFU with a mask in place, generating an image of the fibre slit at the
detector with a gap for every input row. Since the mapping of the input to
the slit reverses every row, to ensure that elements are adjacent both at
the input and output (Fig.~\ref{N4151/ifuinput}), the distance between
illuminated band centres at the detector alternates about the row
separation. By fitting a straight line to the centroids, the magnification
was measured as $2.002 \pm 0.005$ pixels/fibre.

The J-band throughput of the IFU was measured by comparing observations of
the standard stars HD\,84800 and BS\,4069, taken with the CGS4 slit and the
SMIRFS-IFU respectively. Examination of the point-spread function (PSF)
indicates that truncation of the seeing disc by the 4\arcsec\ slit did not
cause significant light loss. The difference in airmass was small, giving an
expected error due to atmospheric extinction of $<1\%$. It was found that
the throughput is 50\%, consistent with the figure of 49\% from the 1997
commissioning run \cite{LeePhD}.

% The airmasses for the 2 stars were 1.17 and 1.10. This means a difference
% in atmospheric extinction of 0.7\%.

\begin{figure}
%\vspace{10cm}
\epsfysize 8 truecm
\epsffile{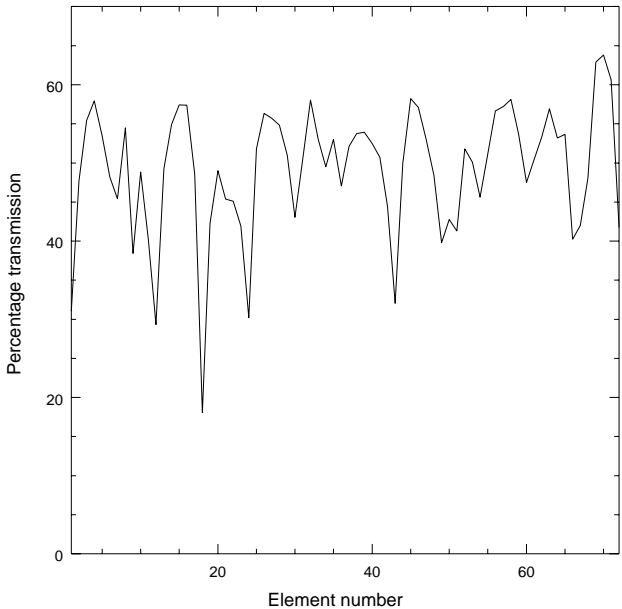}
\caption{Throughput variation along the slit.}
\label{N4151/fibreflat}
\end{figure}

Variations in the fibre transmissions were quantified by extracting a
flat-field spectrum for each element (see \S\ref{ss:N4151/basicred}) and
integrating in wavelength (Fig.~\ref{N4151/fibreflat}). The values
therefore include crosstalk from neighbouring fibres. For Gaussian profiles
with FWHM 1.8 pixels, the contribution from either neighbour is $\sim$10\%
of the total. Ignoring further smoothing due to interpolation and any
correlation between throughputs, the measured and true RMS values are thus
related as follows:
\begin{displaymath}
  \sigma_{\mathrm{data}} = \sqrt{(0.8\sigma_{\mathrm{IFU}})^2
                           +2\times(0.1\sigma_{\mathrm{IFU}})^2}
                         = 0.8\sigma_{\mathrm{IFU}}.
\end{displaymath}
The measured RMS, $\sigma_{\mathrm{data}}$, indicates the signal-to-noise
fluctuation; it is 8\% overall or 6\% excluding the dead fibre and some
vignetted elements. The corresponding values of $\sigma_{\mathrm{IFU}}$,
which characterizes the fibre bundle, are $\sim$10\% and $\sim$7\%
respectively. It is concluded that whilst careful flat-fielding is
important, inherent differences in the signal-to-noise ratio are reasonably
modest.

The point spread function does not vary substantially between fibres. Whilst
a previous estimate \cite{SMIRFS99} indicates that some profiles are up to
20\% broader than others, the PSF is only known to an accuracy of 
$\sim$10\%. Any variations at this level will contribute slightly to the
Nyquist-scale noise in line width maps. The effect on line flux noise is
negligible, since the broadening is a fraction of a pixel. Any
inhomogeneities introduced by the fibres can be smoothed out to some extent
by combining observations with slight spatial offsets.

\section{Data reduction and analysis} \label{s:N4151/dred}

\subsection{Overview}

Data were reduced in IRAF, using several purpose-written scripts and
programs alongside standard tasks. As discussed in Turner (2001),
the best method of dealing with fibre spectra
depends largely on their separation and FWHM at the detector. For the
SMIRFS-IFU, spectra are much more densely packed than in a multi-object
design, precluding the use of existing extraction software. New tasks were
also needed for reconstructing and working with the observed
$x$--$y$--$\lambda$ volumes in the form of datacubes, and for bad pixel
correction. Because the slit maps to the field of view in a complex way,
flexure in the spectrograph cannot be compensated by re-centring the target
(this would be true anyway for off-centre mosaic positions). Observations
must therefore be co-added after conversion to datacubes, rather than in
their raw format. This requires effective bad pixel correction for
integration pairs.

Turner (2001) presents an alternative reduction method for fibre-IFU data,
inspired by the very dense packing of spectra, one per row, in TEIFU
\cite{Teifu00}. With the long CGS4 camera, the raw format of SMIRFS-IFU data
is intermediate between that of TEIFU and multi-object designs. Since fibres
are imaged with a FWHM of 1.8 pixels, the output pattern is almost
critically sampled and is acceptable to interpolate in pixel coordinates.
Moreover, the two pixel fibre separation is large enough that moderate
smoothing by the interpolant will be small in relation to the image
resolution. At the same time, flexure-related errors and undersampling at
the IFU input rule out much gain in precision through TEIFU-style
reduction. Hence, as for most fibre-based instruments, the reduction is
based on extracting 1D spectra.

Finding absolute fibre positions on the detector after flexure is
complicated by vignetting at the slit ends, which prevents direct location
of the first and last elements. With an already limited field of view, it
was decided to rely on the single broken fibre for measuring offsets
between frames, rather than masking off row ends at the input to create
regular gaps. This is usually adequate, if somewhat awkward and
instrument-specific.

With 0.62\arcsec\ elements at the IFU input, undersampling is potentially an
important issue. Inspection of various observations (with and without
SMIRFS) indicates that the seeing during the February 1998 run was
$\sim$0.9\arcsec, including integration over the input microlenses; this
figure is relatively favourable for such coarse sampling. A very basic
simulation of spatial frequency propagation through the instrument shows
that, depending on the alignment of fibres, pixels and image features,
aliased frequencies in the data can have amplitudes up to 10\% of the
mean \cite{Turner01}. However, the Fourier spectrum is only affected
significantly above $\sim$80\% of the Nyquist frequency, so moderate
smoothing during extraction and reconstruction can easily eliminate much of
the contamination. In principle, offset observations can be combined with
phase shifts such that aliasing is removed \cite{Lauer99}, but in this case
the main effect would be to amplify noise because the offsets are not known
or controlled precisely (and the seeing varies). Nevertheless, simply
co-adding datacubes after alignment on a common grid does suppress residual
alias components, which add out of phase like noise. At the end of the
reduction process, aliasing artifacts are not a major source of error.

\subsection{Calibration and extraction} \label{ss:N4151/basicred}

Each observation consists of two integrations, offset by one detector pixel
in the spectral direction to account for bad pixels. These raw frames are
largely obscured by non-uniform dark current. The first step was to
subtract a sky pair from an object pair, removing dark current in the
process. However, this double use means that sky frames cannot be smoothed
spatially to reduce their noise level before subtraction. The result was
then divided by a detector flat-field image, obtained with the calibration
unit before the IFU was installed.

A less accurate dark frame, taken earlier in the night, was also subtracted
from the object separately. This reveals the fibre throughput pattern so
that the dead fibre can be located. Ideally, this would be done using a sky
minus dark image, but flexure of a fraction of a pixel can occur even
between consecutive pointings. This is important for image reconstruction,
but less so for sky subtraction---assuming Gaussian fibre profiles and an
offset of 0.3 pixels, the sky residual due to throughput variations is
likely to be around 3\%, falling below the noise level.

A program was written to combine the pairs of offset, sky subtracted frames
with bad pixel removal. This makes a crude noise estimate based on the RMS
residuals between the two images; any input pixel which differs from the
median of its neighbours by more than a specified number of standard
deviations is excluded. The procedure can sometimes truncate narrow arc lamp
emission lines, whose gradients can be much greater than the noise level,
but is safe for the galaxy spectra (which have had sky lines subtracted
already). Any obvious remaining blemishes were removed manually.

Spectra were straightened with respect to detector rows by applying a small
pre-determined shift to each column. A curvature model was constructed by
measuring the centroid of a suitable image at each wavelength and fitting a
low-order polynomial to the values. Interpolation was performed using a
cubic spline, which is good for dealing with slightly undersampled data.
Inspection of corrected frames verifies that the curvature is practically
constant along the slit and hence independent of flexure. The maximum
difference between columns is 0.7 rows.

Absolute fibre positions were found by summing the straightened object
minus dark image in wavelength and fitting the dead fibre position using
standard IRAF routines. The corresponding sky subtracted image was
resampled so that fibres fell at the mid-points between odd pairs of
rows. These pairs were combined with the \emph{blkavg} task, producing a
stack of extracted fibre spectra (boundary rows were discarded). Sometimes
spatial gradients make the dead fibre difficult to identify, but the
procedure can be repeated if a reconstruction appears distorted.

\begin{figure*}
\epsfysize 9 truecm
\epsffile{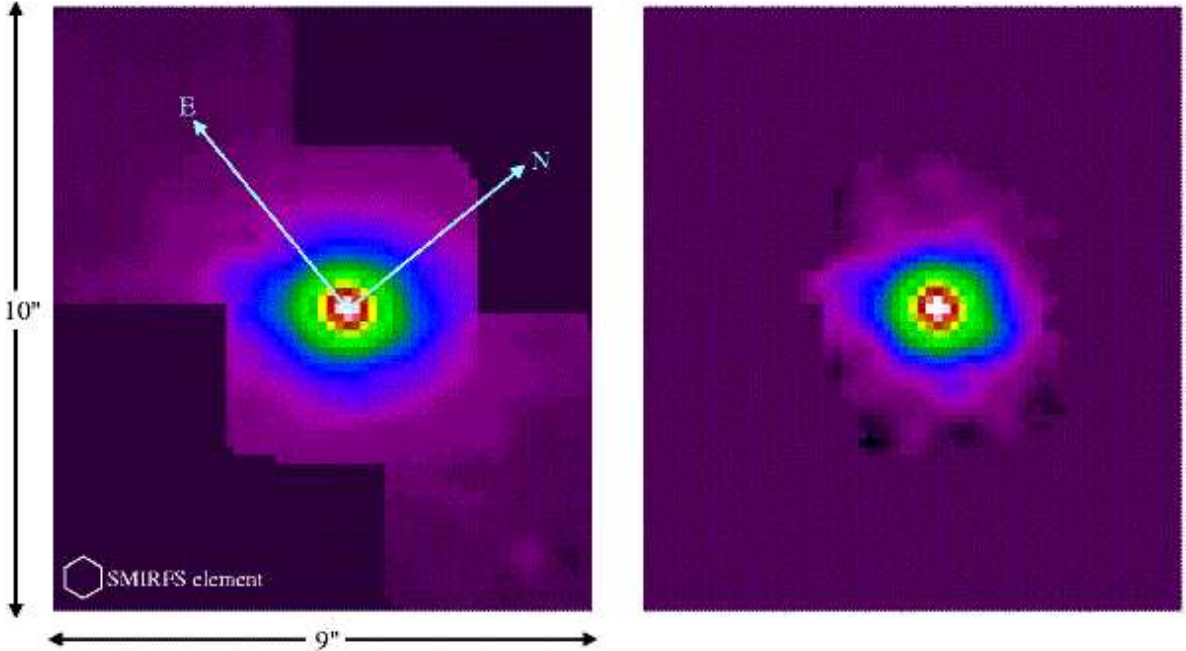}
\caption{(a)~1.25--1.3\um\ image of the nucleus of
NGC\,4151. (b)~Paschen~$\beta$ broad line image. This indicates
the spatial resolution.}
\label{N4151/continpsf}
\end{figure*}

The stack of extracted spectra was divided by a fibre flat, created in the
same way but averaged over its wavelength range. Sky emission lines in the
object minus dark image were used for wavelength calibration. These allow
correction of grating offsets between observations and also provide an
absolute calibration. The spectra from every observation were interpolated
onto a suitable standard grid, truncating the end few pixels which fell
outside the common range. The error in determining the relative offsets is
around 0.2\AA. The error in the linear wavelength fit is 0.7\AA, which is
more than adequate for the present work. Separate solutions were not
determined for the spectra within each observation; variation along the
slit ($\pm0.3$\AA) is smaller than the final random velocity errors in the
mosaic and much smaller than the true gradient in the galaxy.

Finally, the stack was divided by a continuum-normalized stellar spectrum,
to remove telluric absorption features. A \PaB\ line in one of the stars
was interpolated over. Unfortunately, the mosaicing process makes it
difficult to re-observe the standard stars frequently, so there was
sometimes a non-negligible difference in airmass from the galaxy. Although
the main absorption band at 1.269\um\ has been removed effectively in the
final data, there may be some absorption residuals at around 2\% of the
continuum level.

\subsection{Construction of datacubes and mosaicing}

Sets of extracted spectra were interpolated spatially at each wavelength
step, onto a square grid of 0.15\arcsec\ 
pixels.\footnote{We used 
the method of Renka and Cline\cite{Renka83} for convenience.}
The constituent datacubes of each mosaic cycle were combined with
appropriate offsets using the program \emph{mosaic}, which has been included
as a prototype task in the \emph{imspec} package \cite{Turner01}.  This
calculates the smallest common grid from a list of pointings and resamples
the images at each wavelength using a tapered {\it sinc} function. The precise
interpolation method is not important, since the datacube grids are much
finer than the resolution of the data. A mask image is also produced,
recording the number of input pixels contributing at each point.

% The algorithm fits a smooth, continuous surface of quadratic pieces, passing
% through the original samples. Partial derivatives are determined by
% distance-weighted least-squares fitting to nearby points. Similar procedures
% are available freely from Netlib and Starlink on the Internet.  As long as
% the points remain in their original grid, one could also resample each
% dimension in turn using a suitable 1D interpolant, although the present
% method is more general, placing the samples according to a list. The new
% grid is bounded spatially by extremal fibre centres and the images are
% stacked in a datacube with the same wavelength elements as the input
% spectra.

Unfortunately, telescope pointing errors of a fraction of a fibre are
evident. This is partly because a dichroic mirror is used to guide on the
target at visible wavelengths whilst observing in the infrared; the two
images drift apart as atmospheric dispersion changes with the airmass along
the line of sight. Modelling this effect may systematically reduce errors,
but was outside the scope of the current project. There may consequently be
some image distortion at the points of overlap; mosaicing nevertheless
reveals any wider-scale extension of spectral features within the noise
limit of the data. For the NGC\,4151 observations, emission turns out to be
confined mainly to the central field; three mosaic positions with
appreciable signal-to-noise, offset diagonally east and west, were finally
used for mapping.

After constructing each separate mosaic, the resulting datacubes were
combined using \emph{comcubes}, also included in \emph{imspec}. This is
very similar to \emph{mosaic}, but calculates the offsets between datacubes
by locating the centroids of the intensity peaks after integrating in
wavelength. The individual overlap masks are also combined to show the
total number of data points contributing at each output pixel. The final
datacube is divided by the mask, to scale the pixels correctly.

The spectral response of the instrument was determined by fitting piecewise
polynomials to the continua of the standard star observations, dividing the
results by black-body curves appropriate for their spectral types. This was
achieved using the \emph{standard}, \emph{sensfunc} and \emph{calibrate}
tasks in the IRAF \emph{onedspec} package. The response varies at the level
of 5\%, and with atmospheric absorption features already removed it is not
strongly dependent on airmass. The accuracy to which the curve can be
determined is limited mainly by the wings of absorption lines in the stars
and atmosphere, which are difficult to distinguish from modulation in the
continuum. The final datacube was divided by the response function using
\emph{calibrate} directly and the values normalized to a unit mean.

\subsection{Mapping the flux and velocity field}

A velocity map was obtained from the combined mosaic.  Spectra with mean
fluxes below 2.5\% of the central value (about $3\sigma$) were not included,
since line emission is barely detected at the corresponding distance. The
NOAO task \emph{rvidlines} was used to measure a combined \FeII\ and narrow
\PaB\ redshift for each spectrum. This is more reliable for low
signal-to-noise peripheral points than using \FeII\ alone. The results show
that differences in the velocity fields of the two lines are comparable with
the measurement errors and small compared to the line widths. Hence any
deviations introduced by \PaB\ will not affect line strength measurements
significantly. The plain text results from \emph{rvreidlines} were
redirected into a file and converted to a 2D image using \emph{listoim} in
\emph{imspec}.

Line fluxes were measured using the \emph{specmap} task in
\emph{imspec}. This reads the datacube, velocity map and a list of
wavebands for line and continuum estimates, producing a flux map as
output. The measurement bands and linear continuum estimates can be
overplotted on the spectra for verification. Over the wavelength range
observed, there is little pure continuum to provide a suitable reference
for \FeII, but experimenting with different bands shows that the derived
spatial profiles are robust.

An excitation map, showing the strength of \FeII\ relative to \PaB, was
also produced. Since this involves dividing one image by the other, the
result is sensitive to noise where emission is weak. The map was therefore
created using line images from a spatially smoothed version of the
datacube. Rough line width estimates were made by examining individual
spectra with \emph{implot} and \emph{splot}.

\section{Results and comparison with other data}

\begin{figure}
%\vspace{10cm}
\epsfysize 6 truecm
\epsffile{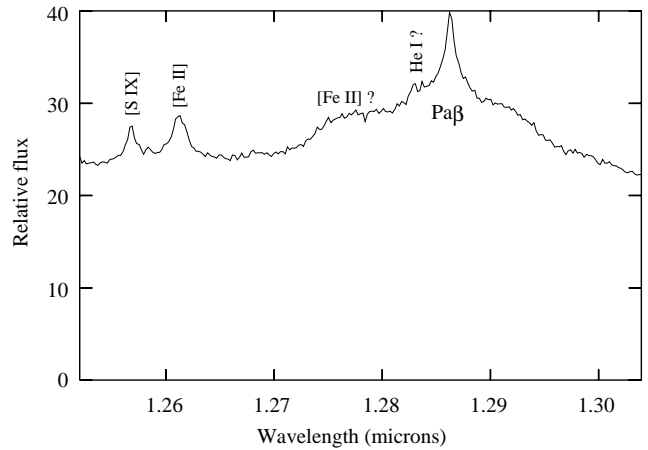}
\caption{Central spectrum of NGC\,4151.}
\label{N4151/censpec}
\end{figure}

\begin{figure}
%\vspace{10cm}
\epsfysize 10 truecm
\epsffile{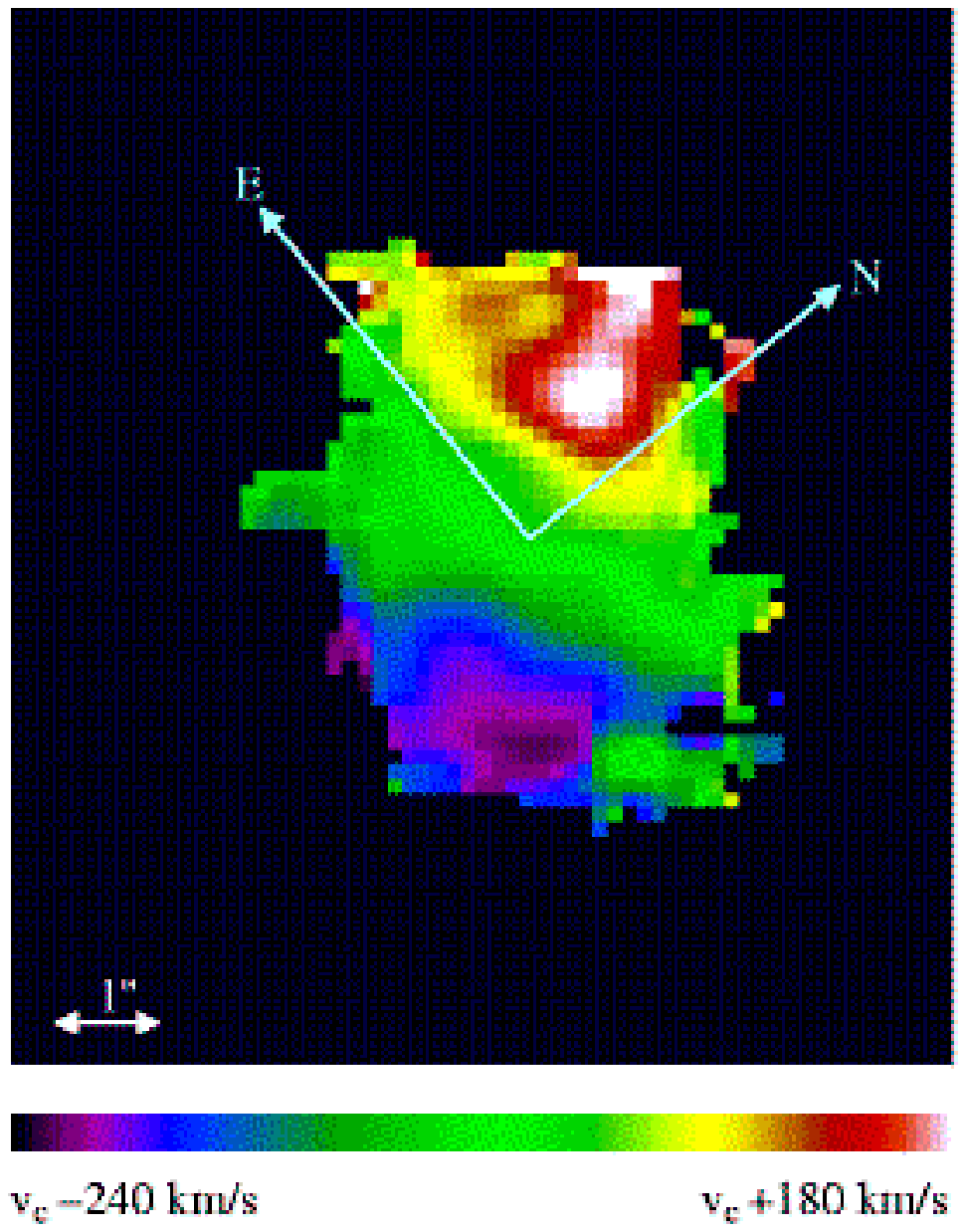}
\caption{Combined \FeII\ and \PaB\ velocity map.}
\label{N4151/velmap}
\end{figure}

A wavelength-integrated image of NGC\,4151 is shown in
Fig.~\ref{N4151/continpsf}(a).  Whilst some horizontal broadening is
caused by the IFU, because spectra overlap, there is also a real NW--SE
extension of the continuum, following the stellar distribution seen in
optical images. Fig.~\ref{N4151/continpsf}(b) is an image of the \PaB\
broad line (including the narrow line at $\sim$6\% of the total), which
gives an indication of the spatial resolution.

A central spectrum is presented in Fig.~\ref{N4151/censpec}; this
includes the broad and narrow \PaB\ components, \FeII\ and \SIX. There may
also be contributions from a weaker 1.2703\um\ \FeII\ feature, superimposed
on the blue wing of the \PaB\ broad line, and \HeI\ at 1.2791\um, to the
left of the \PaB\ narrow line. The small bump to the right of \SIX\ is
probably due to imperfect removal of atmospheric absorption features.

The combined \FeII\ and \PaB\ velocity map (Fig.~\ref{N4151/velmap})
traces the rotation of gas about an axis projected at a position angle of
$\sim$110$^{\circ}$. It is broadly consistent both with previous slit
measurements of the same features \cite{Knop96} and with optical
spectroscopy of the NLR \cite{Mediavilla95,Winge99}. Estimates of
the line widths are complicated by the \PaB\ broad line shape, lack of
clear continuum and modest signal-to-noise ratio off centre. The line
profiles have not been modelled with multiple components, but their overall
deconvolved widths are close to the values found by Knop et al,
who note that the width of \FeII\ relative to \PaB\ suggests a contribution
from shock excitation. At the centre, the FWHM values are approximately
400\,\kms\ for \FeII\ and 250\kms\ for \PaB. Both lines are broader in the
brightest parts of the surrounding extended emission (see below), reaching
$\sim$600\,\kms\ and $\sim$400\,\kms\ respectively. Additional broadening
in the direction of the radio jet is not evident.

%% below added by JRAS, 23/10/01
As an alternative to Knop et al.'s suggestion that the greater velocity
width of the \FeII\ emission is due to a contribution from shock excitation,
we note that our result is consistent with that for a sample of Seyfert 2
galaxies \cite{Veilleux97}. Veilleux et al. cite this as evidence that the
two lines are emitted in different volumes of gas. We also note that Kaiser
et al. (2000) found different components in the velocity field of NGC4151
revealed by [OIII] emission.  They suggested that the low velocity
dispersion component arises from smooth flow in the gravitational field of
the galaxy while the high dispersion component results from an outflowing
wind.  Our result could arise if the \PaB/\FeII\ ratio was higher in the
low dispersion component than in the high dispersion component.
%%%%%%%%%%%%%%%%%%%%%%%

\begin{figure*}
%\vspace{10cm}
\epsfysize 10 truecm
\epsffile{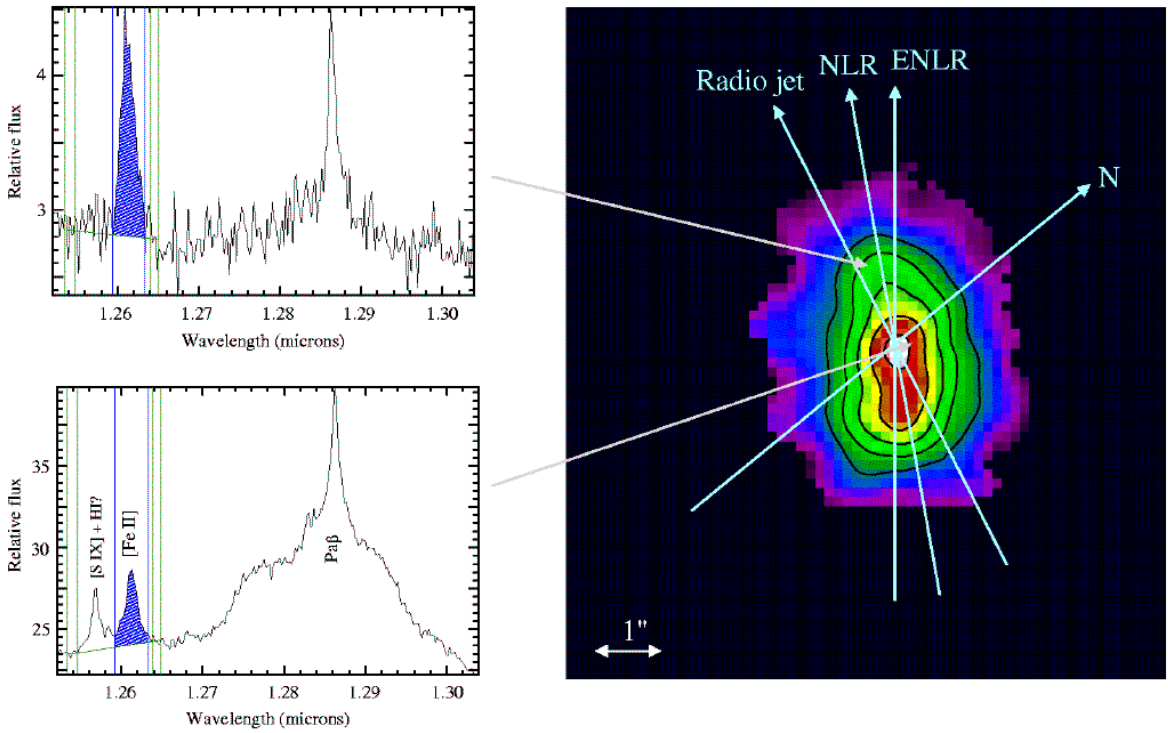}
%%%%% 2nd sentence of caption added by JRAS, 23/10/01
\caption {Map of \FeII\ flux, with measurements of sample spectra. The axes
marked are from Winge et al. (1999) for the ENLR and Kaiser et al. (2000) 
for the NLR and radio emission. } 
\label{N4151/Femap}
\end{figure*}

The spatial distribution of \FeII\ is mapped in Fig.~\ref{N4151/Femap},
alongside examples of individual spectra. The image profile is
extended along both axes of the IFU with an intrinsic RMS of 0.3--0.4\arcsec\  
horizontally and 0.9--1.0\arcsec\ vertically (FWHM 0.8\arcsec\ and
2.2\arcsec\ respectively assuming a gaussian distribution). The
narrowest point is at the intensity peak, as might be expected for a
biconical shape. Although these scales are comparable to the image
resolution, it is clear that the \FeII\ is aligned closer to the optical
line emission than to the radio axis. The best fit to the slope of the
horizontal centroid has a PA of $\sim$54$^{\circ}$, part way between the NLR
and ENLR axes. For the data to fit the radio axis equally well, either end
of the outside ($\sim$25\%) contour would have to be displaced by
$\sim$0.8\arcsec.

\begin{figure*}
%\vspace{10cm}
\epsfysize 10 truecm
\epsffile{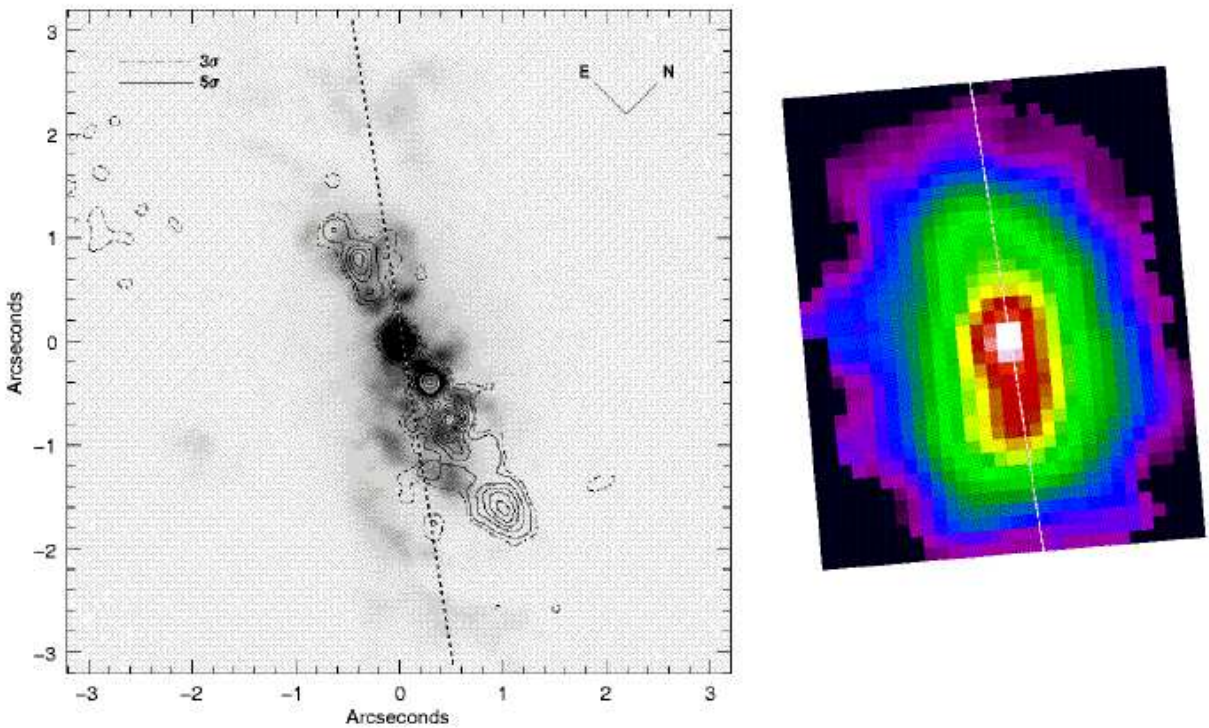}
\caption{ 
(left) HST image of
NGC\,4151 in \OIII\ emission, overlaid with radio contours, from
Kaiser et al. (2000). (right) SMIRFS-IFU image of \FeII\ on the same
scale. Dotted lines mark the \FeII\ axis.
}
\label{N4151/iroptrad}
\end{figure*}

Fig.~\ref{N4151/iroptrad} compares the \FeII\ map directly with a
narrow-band HST image of \OIII\ line emission \cite{Kaiser00} and a contour
map of the radio jet at 8\,MHz \cite{Mundell95}.  It can be seen that the
region over which \FeII\ is detected in the SMIRFS data corresponds to the
extent of the narrow line region. If there were a strong correspondence with
the radio jet, one might expect a more elongated distribution, resolving
some knot structure at $77^{\circ}$, where the outflow interacts with
gas. Although detection of the larger-scale ENLR would have provided a
better separation, the spatial association of \FeII\ with photoionized gas
is more convincing than with shocking by radio plasma. The F502N filter used
for imaging \OIII\ covers almost all of the NLR velocity range
\cite{Hutchings99,Kaiser00}, so represents the total emission well. It is
perhaps interesting to note that \FeII\ appears better correlated with the
positions of low velocity dispersion clouds, plotted in \cite{Kaiser00},
than the NLR as a whole. 
%% below added 24/10/01
However, this is at variance with our previous
suggestion that the greater velocity width of the \FeII\ line compared with
\PaB\ suggests an association with the high dispersion component.
%%%%%%%%%%%%%%%%%%%%%%%

\begin{figure*}
%\vspace{10cm}
\epsfysize 11 truecm
\epsffile{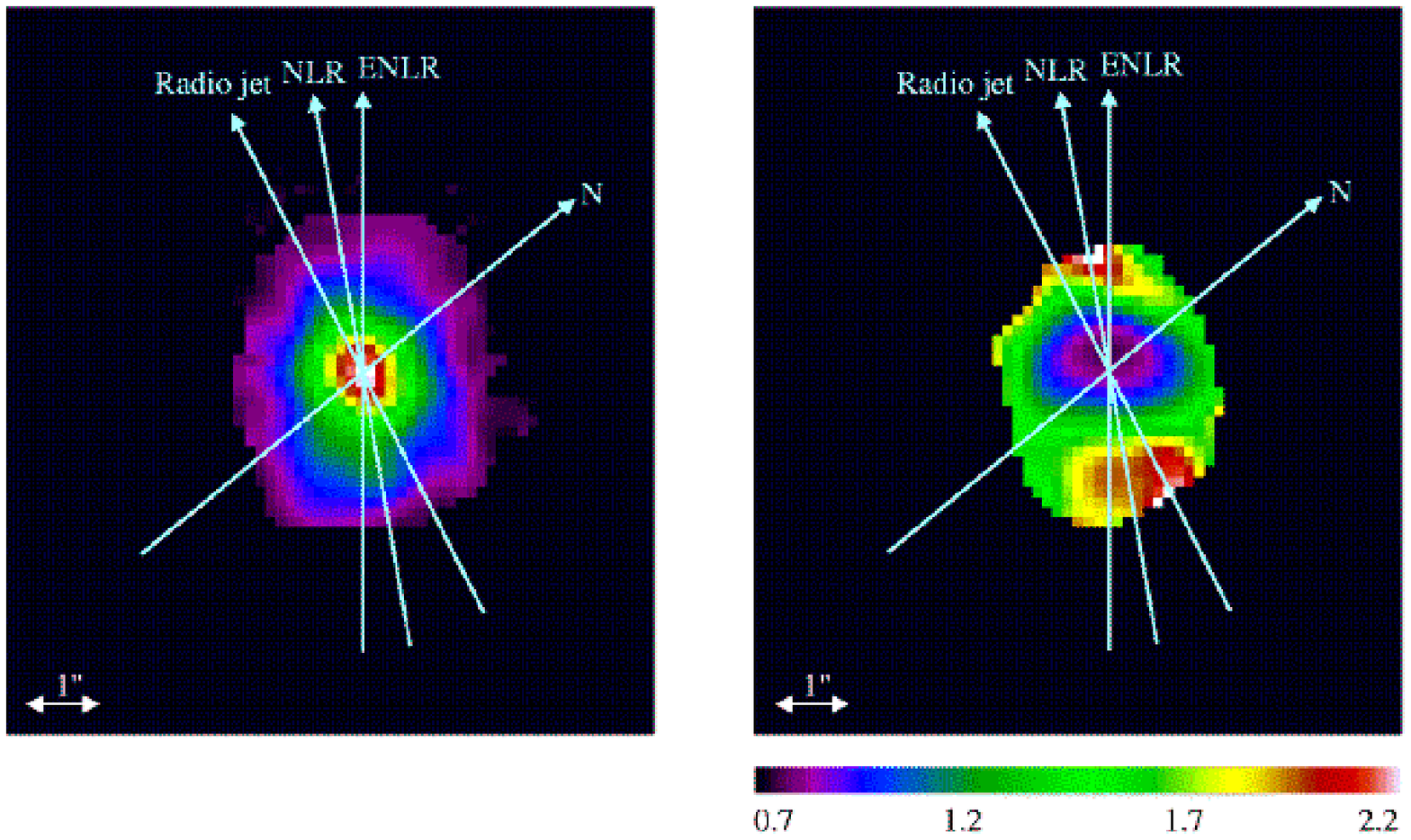}
\caption{(left) \PaB\ emission (right) Ratio of \FeII/\PaB.}
\label{N4151/linerat}
\end{figure*}

The \PaB\ narrow line is slightly extended in both directions, and has a
faint wing out to $\sim$1.5\arcsec\ South-West of the centre, where the NLR
is brightest. Fig.~\ref{N4151/linerat} shows both the \PaB\ flux and
\FeII/\PaB\ ratio. Since \PaB\ falls off fairly rapidly from the centre, the
ratio is highly sensitive to noise artifacts. The datacube was therefore
smoothed with a 0.9\arcsec\ wide Gaussian kernel before producing the
excitation map, preserving minimal resolution of spatial structure. The line
ratio is $\sim$0.7 at the centre and $\geq 2$ along the radio jet at the
edges. A dip near the centre is to be expected for photoionization, since
the strong incident flux creates large fully ionized zones, favouring \PaB\
over \FeII\ \cite{Mouri00}.  The high areas could reflect an enhancement of
\FeII\ emission by shocks, but the line images are more indicative of a drop
in \PaB\ at those points, with respect to \OIII, than of strong \FeII. To
the South-West, a few highly-ionized clouds may just happen to give stronger
\PaB\ away from the radio axis, whilst to the North-East, the NLR is angled
slightly more towards the radio jet anyway.

\section{Conclusions}

The distribution of \FeII\ emission at the centre of NGC\,4151 has been
mapped in two dimensions, using the innovative SMIRFS-IFU. Although the
line is only detected over a few spatial resolution elements, it is evident
that its image profile is better aligned with the visible narrow line
region than with the radio jet. The study therefore suggests that
\FeII\ arises primarily through photoionization of gas, by collimated
X-rays from the Seyfert nucleus. The mean velocity field is consistent with
this interpretation, indicating that \FeII, \PaB\ and the NLR all have
similar bulk kinematics. Likewise, the variation in \FeII/\PaB\ is
compatible with photoionization dominating. As suggested by previous
authors, shock excitation by outflowing radio plasma may make a secondary
contribution to the emission. However, the evidence for this comes mainly
from line widths, whereas tentative measurements suggest, if anything, that
\FeII\ is broader along the NLR than the radio jet.

Despite the small field and prototypical nature of the IFU, these results
demonstrate the utility of integral field
spectroscopy in investigating the properties of active galaxies. Without a
two dimensional field, it would have been very difficult to determine the
spatial orientation of the line emission. Narrow-band imaging can also
fulfil this requirement, given appropriate filters, but follow-up
spectroscopy is then required to form a complete picture. The experience
gained with the SMIRFS-IFU and associated data analysis has paved the way
for subsequent projects, both technical and observational.

Reduction of the IFU data has highlighted a few important observational
issues. In particular, location of spectra on the detector is difficult,
and it is recommended that more than one reference gap between fibres
should be available in future work.
%If possible, the accuracy of mosaic offsets might be improved by tilting
%the secondary mirror instead of moving the telescope. Finally,
With a number of sources of error and inhomogeneity, the availability
of multiple observations at each position has been very valuable.

Whilst the results presented here support the argument for \FeII\
originating in photoionized gas, there is no doubt that observations at
high spatial resolution, as well as of the wider ENLR, will provide further
insight into the nature of NGC\,4151 and other Seyfert galaxies. With a
bright, point-like nucleus, NGC\,4151 is an ideal target for J-band
observations with adaptive optics. This would provide a useful counterpart
to the Hubble Space Telescope observations of \OIII\ emission, which have
revealed the cloud structure and detailed kinematics of the NLR. In
conjunction with recent models of photoionized and shock excited gas,
comparison of optical and infrared lines in individual clouds would provide
important information about the physical conditions and kinematic structure
of the NLR. The misalignment with the radio jet will again help determine
its influence on the ambient gas.

\section*{Acknowledgements}

The authors thank George Dodsworth and Ian Lewis for their work on the
design and construction of the SMIRFS IFU and Ray Sharples, Ian Parry and
Reynier Peletier for their valuable contributions. The United Kingdom
Infrared Telescope is operated by the Joint Astronomy Centre on behalf of
the U.K. Particle Physics and Astronomy Research Council.

\end{document}